\documentclass[a4paper,11pt]{article}
\pdfoutput=1

\usepackage{jheppub} 

\usepackage[T1]{fontenc} 

\usepackage{graphicx, epsfig}
\usepackage{color}
\usepackage{mathrsfs}
\usepackage{bm}
\usepackage{amsmath,amssymb}
\usepackage{mathrsfs}
\usepackage[normalem]{ulem}
\usepackage{mathtools}
\usepackage[x11names]{xcolor}
\usepackage{caption}
\usepackage{subcaption}
\usepackage{multirow}
\usepackage{float}
\usepackage{hyperref}

\definecolor{fashionfuchsia}{rgb}{0.96, 0.0, 0.63}
\colorlet{no_so_fashion_purple}{blue!50!red}

\newcommand{\be}{\begin{equation}}
	\newcommand{\ee}{\end{equation}}
\newcommand{\ba}{\begin{eqnarray}}
	\newcommand{\ea}{\end{eqnarray}}

\newcommand{\nn}{\nonumber}

\newcommand{\bfB}{{\bf B}}
\newcommand{\bfA}{{\bf A}}

\newcommand{\m}{{m}}
\newcommand{\Phimm}{\Phi_{m {\bar m}}}

\newcommand{\Sinm}{\sin ( \theta_m/2 ) }
\newcommand{\Sinmbar}{\sin ( \theta_{\bar m}/2 ) }
\newcommand{\Cosm}{\cos ( \theta_m/2 ) }
\newcommand{\Cosmbar}{\cos ( \theta_{\bar m}/2 ) }

\newcommand{\minphi}{{\rm Min}[|\Phi|]}

\title{Annihilation of electroweak dumbbells}

\author{Teerthal Patel$^{*}$}
\author{Tanmay Vachaspati$^{*,\dag}$}

\affiliation{
$^*$Physics Department, Arizona State University, Tempe,  Arizona 85287, USA.\\
$^\dag$Theoretical Physics Department, CERN, 1211 Geneva 23, Switzerland.
}

\emailAdd{tpatel28@asu.edu}
\emailAdd{tvachasp@asu.edu}

\abstract{
We study the annihilation of electroweak dumbbells and the dependence of their dynamics 
on initial dumbbell length and twist. Untwisted dumbbells decay rapidly while maximally
twisted dumbbells collapse to form a compact sphaleron-like object, before
decaying into radiation.
The decay products of a dumbbell include electromagnetic magnetic fields
with energy that is a few percent of the initial energy. The magnetic field from the
decay of twisted dumbbells carries magnetic helicity with magnitude that depends
on the twist, and handedness that depends on the decay pathway.
}

\begin{document}
{\hfill CERN-TH-2023-200} 

\maketitle
\flushbottom

\section{Introduction}
\label{sec:intro}

The ``electroweak dumbbell'' consists of a magnetic monopole and an antimonopole of the
standard electroweak model connected by a string 
made of $Z$ magnetic field~\cite{Nambu:1977ag,Achucarro:1999it}.
The existence of such non-perturbative field configurations in the electroweak model is of 
great interest as they can provide the first evidence for (confined) magnetic monopoles. In a 
cosmological context, dumbbells can source large-scale magnetic fields which can seed galactic 
magnetic fields and play an important role in the propagation of cosmic rays~\cite{Vachaspati:2020blt}. 

Electroweak dumbbells are often regarded as magnetic dipoles, with the magnetic field strength falling 
off as  $1/r^3$ with the distance $r$ from the dipole. However the situation is richer: there is a one-parameter 
set of electroweak dumbbell configurations \cite{VachaspaticandField}, all describing a confined 
monopole-antimonopole pair, but with additional structure called the ``twist''. 
In our previous work \cite{Patel:2023sfm}, we have shown that the magnetic field strength of the 
maximally twisted dumbbell falls off asymptotically as $\cos\theta/r^2$ (in spherical coordinates), 
a gross departure from the usual dipolar magnetic field. 
The twisted dumbbell configuration was proposed in \cite{VachaspaticandField} and is closely 
related to the 
electroweak sphaleron~\cite{Manton:1983nd,Klinkhamer:1984di,PhysRevD.46.3587,Akiba:1989xu}. 

The dynamics of electroweak sphalerons and dumbbells are of particular interest in the efforts 
towards detecting them in experiments.
In view of ongoing experiments like the Monopole and Exotics Detector (MoEDAL) at the 
Large Hadron Collider (LHC)~\cite{Acharya_2022}, Ref.~\cite{Ho_2020} recently studied
the production of the electroweak sphaleron in the presence of strong magnetic fields that
arise during heavy ion collisions.
In the cosmological context, simulations of the dynamical decay of electroweak sphalerons have 
been conducted to study baryogenesis and magnetogenesis \cite{Copi_2008,Chu_2011}.
Thus, there have been several efforts to numerically resolve the configuration and dynamics of 
electroweak sphalerons.
The relevance of dynamics of electroweak dumbbells was first alluded to by 
Nambu~\cite{Nambu:1977ag} wherein he showed that electroweak dumbbells could be stabilized by 
rotation, potentially making them long lived enough to have significant implications in experimental 
searches. This, along with a lack of detailed study of the dynamics of electroweak dumbbells, has motivated 
our investigation into the dumbbell configurations and their dynamics.

In this article, we investigate the dynamical evolution of an initially stationary dumbbell configuration 
for a range of twists and initial dumbbell lengths.
The initial condition is found by numerically relaxing a ``guess'' dumbbell configuration under the 
constraint that locations of the monopole and antimonopole remain fixed,
according to the method outlined in our previous work \cite{Patel:2023sfm}. 
The main quantities of interest that we analyze are the dumbbell lifetimes and the magnetic field
produced during the decay.
The untwisted dumbbells are found to be unstable, with the monopoles immediately undergoing 
annihilation as expected.
Twisted dumbbells, on the other hand, lead to the creation of an intermediate sphaleron configuration 
after the initial collapse, and subsequently decay into helical magnetic fields with relatively stronger
field strength.

The article begins with a description of the model and our initial configuration in Section~\ref{sec:model}. 
The numerical simulation setup is described in Section~\ref{sec:numerical simulation}. Our
results for the lifetimes and the relic magnetic fields are given in Section~\ref{sec:results}, and
we discuss our conclusions in Section~\ref{conclusions}.
		
\section{Model}
\label{sec:model}

The Lagrangian for the bosonic sector of the electroweak theory is given by
\be
\mathcal{L} = -\frac{1}{4}W^a_{\mu\nu}W^{a\, \mu\nu} - \frac{1}{4}Y_{\mu\nu}Y^{\mu\nu}
+ |D_\mu\Phi|^2 -\lambda(|\Phi|^2-\eta^2)^2\, ,
\ee
where
\begin{equation}
D_\mu \equiv \partial_\mu - \frac{i}{2}g\sigma^aW^a_\mu - \frac{i}{2}g'Y_\mu\, .
\label{covD}
\end{equation}
Here, $\Phi$ is the Higgs doublet, $W^a_\mu$ are the SU(2)-valued gauge fields with $a=1,2,3$ and, 
$Y_\mu$ is the U(1) hypercharge gauge field. In addition, $\sigma^a$ are the Pauli spin matrices
with ${\rm Tr}(\sigma^a \sigma^b) = 2 \delta_{ab}$, 
and the experimentally measured values of the parameters that we adopt from \cite{Particledatagroup2022}  
are $g=0.65$, $\sin^2\theta_w = 0.22$, $g'=g\tan\theta_w=0.35$, $\lambda=0.129$ and $\eta=174\, {\text{GeV}}$.
	
The classical Euler-Lagrange equations of motion for the model are given by
\ba
&&
D_\mu D^\mu \Phi + 2\lambda(|\Phi|^2  - \eta^2)\Phi = 0 \label{EL phi}\\
&&
\partial_\mu Y^{\mu\nu} = g'\, {\rm Im}[\Phi^{\dagger} ( D^\nu\Phi ) ]\label{EL B}\\
&&
\partial_\mu W^{a\mu\nu} + g\epsilon^{abc} W^b_\mu W^{c\mu\nu} = 
g\, {\rm Im}[ \Phi^{\dagger} \sigma^a ( D^\nu \Phi )]
\label{EL W}
\ea
where the gauge field strengths are given by
\ba
W^{a}_{\mu\nu} &=& \partial_{\mu} W^a_\nu - \partial_{\nu} W^a_\mu + g \epsilon^{abc} W^b_\mu W^c_\nu \\
Y_{\mu\nu} &=& \partial_{\mu} Y_\nu - \partial_\nu Y_{\mu} \, .
\ea

Electroweak symmetry breaking results in three massive gauge fields, namely the two charged
$W^\pm$ bosons and $Z_\mu$,
and one massless gauge field, $A_\mu$, that is the electromagnetic gauge field. We define
\ba
Z_\mu &\equiv& \cos\theta_w n^a W^a_\mu + \sin \theta_w Y_\mu, \label{Z def unitary gauge}  \\
A_\mu &\equiv& - \sin\theta_w n^a W^a_\mu + \cos \theta_w Y_\mu\,,
\label{A def unitary gauge}
\ea
where
\begin{equation}
	\label{Higgs n vector}
	n^a \equiv \frac{\Phi^{\dagger}\sigma^a\Phi}{|\Phi|^2} \,
\end{equation}
are components of a unit three vector $\hat{n}$.
The weak mixing angle, $\theta_w$ is given by $\tan{\theta_w} = g'/g$, the $Z$ coupling is defined as 
$g_Z \equiv \sqrt{g^2+g'^2}$, and the electric charge
is given by $e = g_Z \sin\theta_w \cos \theta_w$.
The Higgs, $Z$ and $W$ boson masses are given by 
$m_H \equiv 2\sqrt{\lambda}\eta =125\, {\text{GeV}}$, 
$m_Z \equiv g_Z\eta/\sqrt{2} =91\, {\text{GeV}}$ 
and $m_W \equiv g\eta/\sqrt{2}=80\, {\text{GeV}}$, respectively.

\subsection{Initial dumbbell configuration}
\label{subsec:initial-dumbbell-configuration}

\begin{figure}
\centering
\includegraphics[width=0.4\textwidth]{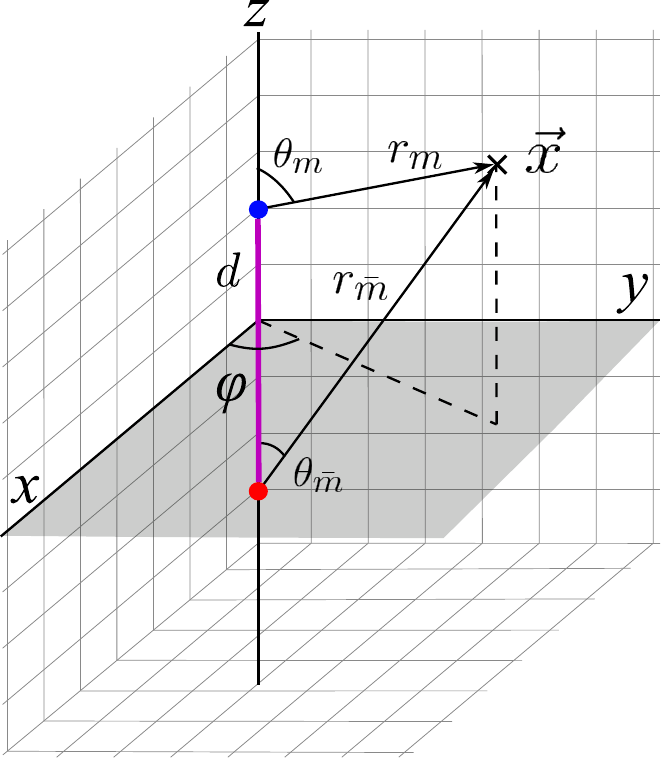}
\caption{The location of the monopole (blue dot), antimonopole (red dot) and the $Z$-string (purple line) in a cartesian grid.}
\label{fig:illustration}
\end{figure}

We construct a suitable initial configuration for dumbbell dynamics by first choosing 
a ``guess'' field configuration that contains a monopole and antimonopole separated 
by a distance $2d$ and with relative twist $\gamma$~\cite{Achucarro:1999it}. In the
asymptotic region,
\begin{equation}\label{Phimmbar}
{{ {{\hat \Phi}_{\m{\bar \m}}}}(\gamma)} = \left( 
\begin{array}{c}
\Sinm \Sinmbar e^{i\gamma} + \Cosm\Cosmbar
\\
\Sinm \Cosmbar e^{i\phi} - \Cosm \Sinmbar e^{i(\phi - \gamma)}
\end{array}
\right)\, .
\end{equation}
The monopole and antimonpole are located along the z-axis at $z=\pm d$, with $ \theta_\m $ 
and $\theta_{\bar{\m}} $ being the spherical polar angles as measured from the monopole and 
antimonpole, respectively; $\phi $ is the azimuthal angle. The coordinate system is illustrated in 
Figure~\ref{fig:illustration}. In the limit $ \theta_{\bar{\m}}\to 0 $, \eqref{Phimmbar} reduces
to the monopole configuration, and in the limit $ \theta_\m \to 0 $ to an antimonopole.
In between the monopole and antimonopole, $\theta_\m \to \pi$ and $\theta_{\bar{\m}}\to 0$,
the configuration is that of a $Z-$string.	

The asymptotic gauge field configurations are obtained by setting the covariant derivative of the 
Higgs field to vanish,
\ba
\label{MMBar gauge fields}
gW^a_\mu &=& -\epsilon^{abc}n^{b}{\partial_\mu n^c} 
+ i \cos^2\theta_w n^a(\Phi^{\dagger}\partial_{\mu}\Phi - \partial_{\mu}\Phi^{\dagger}\Phi) \\
g'Y_\mu &=& -i\sin^2\theta_w (\Phi^{\dagger}\partial_{\mu}\Phi - \partial_{\mu}\Phi^{\dagger}\Phi)
\ea
To correctly account for the spatial dependence of the Higgs field around the 
monopole-antimonopole pair, we attach spatial profiles. 
Including this in the ansatz, the initial monpole-antimonopole scalar field configuration is given by
\begin{equation}
\label{general mmbar scalar config}
{{\Phimm}} = k({\vec{r}}\, )h(r_m)h(r_{{\bar{\m}}}) { {{\hat \Phi}_{\m{\bar \m}}}}\, ,
\end{equation}
where $ r_\m $ and $ r_{\bar{\m}} $ are the radial coordinates centered on the monopole and 
antimonpole, respectively, given by
\begin{equation}
r_m = |{\mathbf{x}} - \mathbf{x}_\m|, \,\,\,\, r_{\bar m} = |\mathbf{x} - \mathbf{x}_{\bar m} |\, .
\end{equation}
The function $ k({\vec{r}}\, ) $ is the Z-string profile.	
Similar to the Higgs, we include spatial profiles for the gauge fields as	
\ba
\label{Gauge field with radial profile}
gW^a_\mu &=& l({\vec{r}}\, )j(r_\m)j(r_{\bar{\m}})[-\epsilon^{abc}n^{b}{\partial_\mu n^c} 
+ i \cos^2\theta_w n^a({\hat \Phi}^{\dagger}\partial_{\mu}{\hat \Phi} 
- \partial_{\mu}{\hat \Phi}^{\dagger}{\hat \Phi})] \\
g'Y_\mu &=& l({\vec{r}}\, )j(r_\m)j(r_{\bar{\m}})[-i\sin^2\theta_w ({\hat \Phi}^{\dagger}\partial_{\mu}{\hat \Phi} - \partial_{\mu}{\hat \Phi}^{\dagger}{\hat \Phi})]
\ea
We have previously used numerical relaxation to solve for the Higgs and gauge field 
profile of a static dumbbell in with the constraint that the topology of the
monopole and antimonopole remain fixed during the relaxation process~\cite{Patel:2023sfm}.
We use the same procedure to find the initial field configurations that we will then
evolve to study dumbbell dynamics.
	
\section{Numerical simulation}
\label{sec:numerical simulation}

We have used a numerical relativity technique adapted from \cite{2010nure.book.....B}, 
and previously used in \cite{Vachaspati:2015ahr} to study monopole-antimonopole 
scattering in the SO(3) model.
Adopting the temporal gauge for convenience in numerical implementation, $W^a_0 = Y_0 = 0$,
the evolution equations \eqref{EL phi}-\eqref{EL W} can be written as
\ba
\partial_0^2\Phi&=& D_iD_i\Phi -2\lambda (|\Phi|^2-\eta^2)\Phi	
\label{Phieq} \\
\partial_0^2 Y_i&=& -\partial_jY_{ij} + g'\, {\rm Im}[\Phi^{\dagger} ( D_i\Phi ) ]
\label{Yeq} \\
\partial_0^2 W^a_i&=&	-\partial_jW^a_{ij} - g\epsilon^{abc}W^b_jW^c_{ij}
+g{\rm Im}[ \Phi^{\dagger} \sigma^a ( D_i \Phi )]\,. 
\label{Weq}
\ea

Straightforward discretization of the evolution equations leads to numerical instabilities.
To control the instabilities, we introduce ``Gauss constraint variables'' $\Xi=\partial_iY_i$ and 
$\Gamma^a=\partial_i W^a_i$, with their respective evolution equations
\ba
\partial_0\Xi&=& \partial_0Y_{0i}-g_p^2\{\partial_0Y_{0i}-g'\, {\rm Im}[\Phi^{\dagger} ( \partial_0\Phi ) ]\}\\
\partial_0\Gamma^a&=&\partial_0W^a_{0i} - 
g_p^2\{\partial_0W^a_{0i}+g\epsilon^{abc}W^b_i\partial_0W^c_{i}
-g{\rm Im}[ \Phi^{\dagger} \sigma^a ( \partial_0 \Phi )]\}\, .
\ea
Here, we have introduced the numerical stability parameter $g_p^2$. These equations
reduce to the Gauss constraints in the continuum, regardless of the choice of $g_p^2$. 
However, once the system of equations is discretized for numerical evolution, the term 
in the curly brackets do not always vanish, and a non-zero value of $g_p^2$ ensures 
numerical stability as outlined in \cite{2010nure.book.....B}.
The equations in \eqref{Phieq}-\eqref{Weq} are now written with the Gauss constraint variables
as
\ba
\partial_0^2\Phi &=& 
\nabla^2\Phi  
- i\frac{g}{2}\sigma^aW^a_i\partial_{i}\Phi 
- i\frac{g'}{2}Y_i \partial_{i}\Phi 
- i\frac{g}{2}\sigma^a\Gamma^a\Phi 
- i\frac{g'}{2}\Xi\Phi \nn \\
&& \hskip 0.5 cm 
-i\frac{g}{2}\sigma^aW^a_i(D_i\Phi) 
- i\frac{g'}{2}Y_i(D_i\Phi) 
- 2\lambda(|\Phi|^2 - \eta^2)^2\Phi
\label{Phi-EOM-Gaussvariables}\\
\partial_0^2Y_i &=& 
\nabla^2Y_i 
- \partial_i\Xi 
+ g'{\rm Im}[\Phi^\dagger\sigma^a(D_i\Phi)]
\label{Y-EOM-Gaussvariables} \\
\partial ^2_0 W^a_i &=& 
\nabla^2 W^a_{i} 
+ \partial_i\Gamma^a 
- g\epsilon^{abc} (\partial_kW^b_i)W^c_k 
- g\epsilon^{abc}W^b_i\Gamma^c \nn \\
&& \hskip 0.5 cm 
- g\epsilon^{abc} W^b_k W^c_{ik} 
+ g {\rm Im}[\Phi^\dagger\sigma^a(D_i\Phi)]
\label{W-EOM-Gaussvariables}
\ea

Our simulations are conducted on a regular cubic lattice with Dirichlet boundary conditions 
and the fields are evolved in time using the explicit Crank-Nicholson method with two 
iterations \cite{Teukolsky:1999rm}. 
We adopted phenomenological values for the electroweak model as given in Sec.~\ref{sec:model}. 
We choose to work in units of $\eta$ and set its numerical value to $ \eta=1$ in our simulations.
Then the Higgs mass of 125~GeV in these units is given by 
$ 2\eta\sqrt{\lambda} = 2\sqrt{\lambda} = 0.72$.
Similarly, the $W$ boson mass is $ \m_W = g\eta/\sqrt{2} = 0.46$.
For most of our runs, we use a lattice of size $400^3$ 
with lattice spacing $ \delta = 0.25 \eta^{-1} =0.25$, and time step $dt=\delta/8$. 
The Compton radius of the $W$ boson is $m_W^{-1} = 2.17$. This is also approximately
the radius of the monopole and string in the dumbbell, 
implying that their profiles have a resolution of roughly 10 grid points in our setup.

The Higgs field vanishes at the centers of the monopole, antimonopole and $Z-$string.
Since the initial dumbbell profile involves delicate cancellation of zeros on the dumbbell,
we offset the dumbbell location away from the $z-$axis by half a lattice spacing. That is,
the monopole and antimonopole are at the coordinates $ x=y=\delta/2, z=\pm(d+\delta/2)$,
while the $Z-$string is located at $ x=y=\delta/2$ and parallel to the $z-$axis.

To ensure that the Dirichlet boundary conditions do not significantly affect the dynamics 
of the annihilating dumbbell, we only consider initial dumbbell lengths $2d$ that are
sufficiently smaller than the lattice size. 
The maximum separation considered in our simulations was slightly less than half the 
lattice width and we ensure that the Higgs field is close to the vacuum expectation value 
near the boundaries $|\Phi | \approx 1$. 
We run our simulations for $\sim(5-10)\times T_c$, where $T_c$ is 
the dumbbell lifetime, to study the relic energies in the different fields. 
We ensure that there are no significant effects on the dumbbell dynamics due to reflections 
from the Dirichlet boundary conditions in the time span considered here.
We have tested this by varying the lattice spacing $\delta$ and the total simulation box size.
Thus we ensure that our choice of numerical parameters and boundary conditions do not 
affect the dumbbell dynamics.

\section{Results}
\label{sec:results}	

\begin{figure}	
\includegraphics[width=\textwidth]{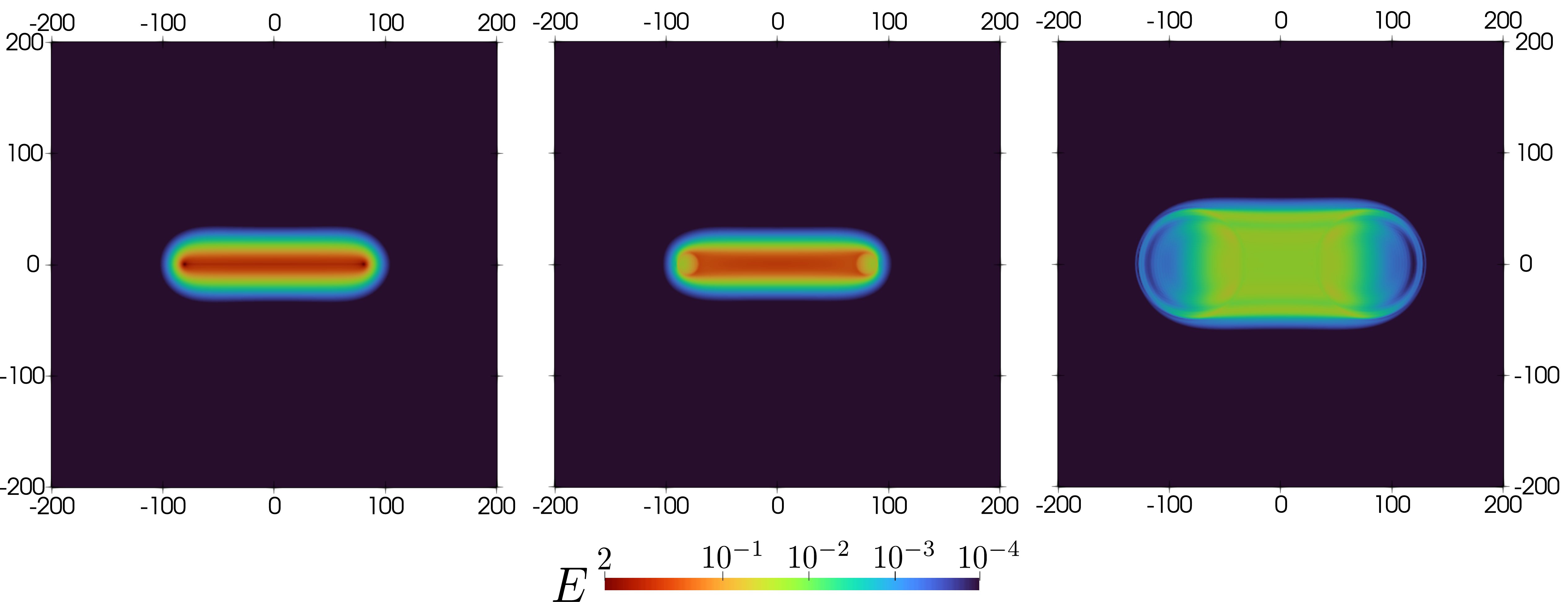}
\caption{The snapshots of energy density in $xz$-plane for $\gamma=0$ are shown for times $0dt$, $1000dt$ and $5000dt$ in the left, middle and right panels, respectively.
The colors represent the energy density and the corresponding values are given by the scale, in units of $\eta^4$.
Here, the simulation parameters are $dt=\delta/100$, with $\delta=0.25\eta^{-1}$, and the initial dumbbell length is $2d=160\delta$.
The horizontal ($z$) and vertical ($x$) axis are in lattice units.
}
\label{fig:energy-slices-twist-0}
\end{figure}

\begin{figure}	
\includegraphics[width=\textwidth]{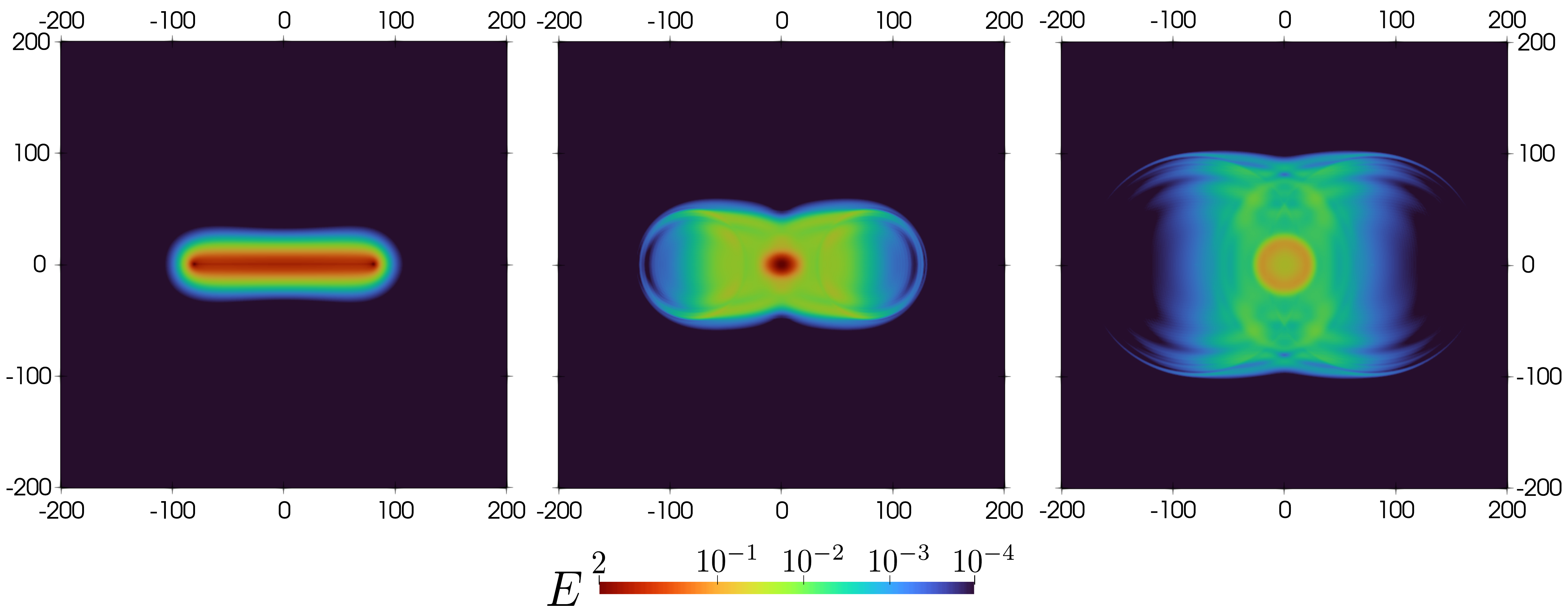}
\caption{The energy density snapshots, similar to Figure~\ref{fig:energy-slices-twist-0}, for the twisted case ($\gamma=\pi$) and times $0dt$, $5000dt$ and $10000dt$ in the left, middle and right panels, respectively.
}
\label{fig:energy-slices-twist-pi}
\end{figure}

We ran the simulation for a range of initial dumbbell lengths and twists.
We show several snapshots of the energy density in the $xz$-plane for the untwisted 
case ($\gamma=0$) in Figure~\ref{fig:energy-slices-twist-0}, and for the twisted case ($\gamma=\pi$) 
in Figure~\ref{fig:energy-slices-twist-pi}. As can be inferred from these slices, the untwisted dumbbell 
promptly undergoes annihilation. However, the twisted dumbbell forms an intermediate object that 
appears as an over-density in the center, before eventually decaying away. We will discuss the relevance 
of this object (most likely the electroweak sphaleron, as discussed in 
Ref.~\cite{VachaspaticandField}) in the context of the dumbbell lifetimes and the 
relic magnetic energies in the following sections.

\subsection{Estimating lifetimes}
\label{subsec:lifteimeest}

\begin{figure}	
\includegraphics[width=0.5\textwidth]{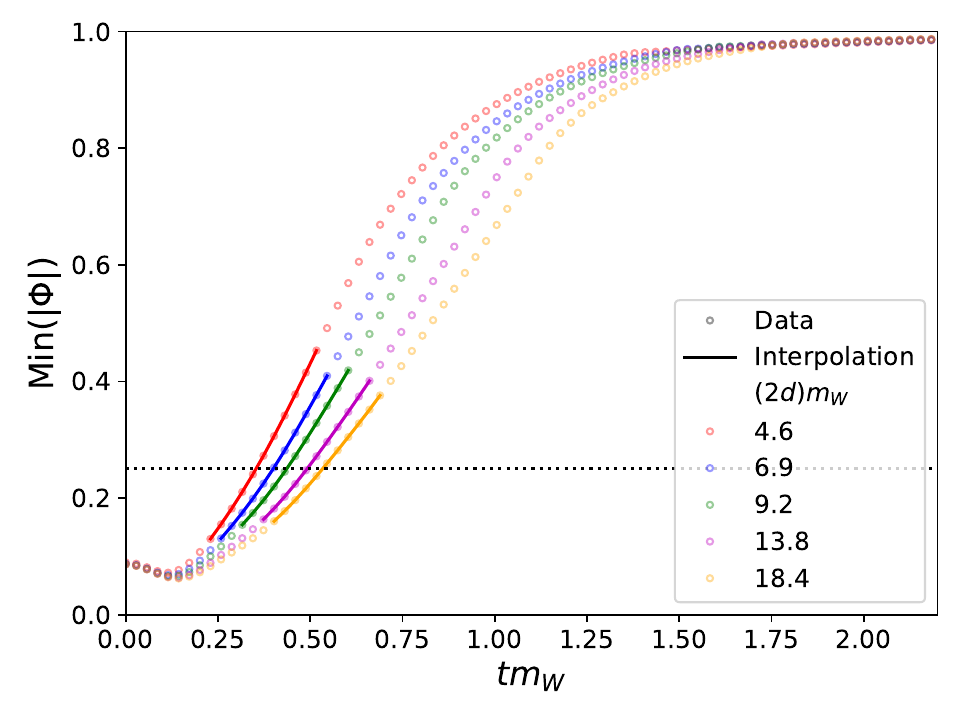}
\includegraphics[width=0.5\textwidth]{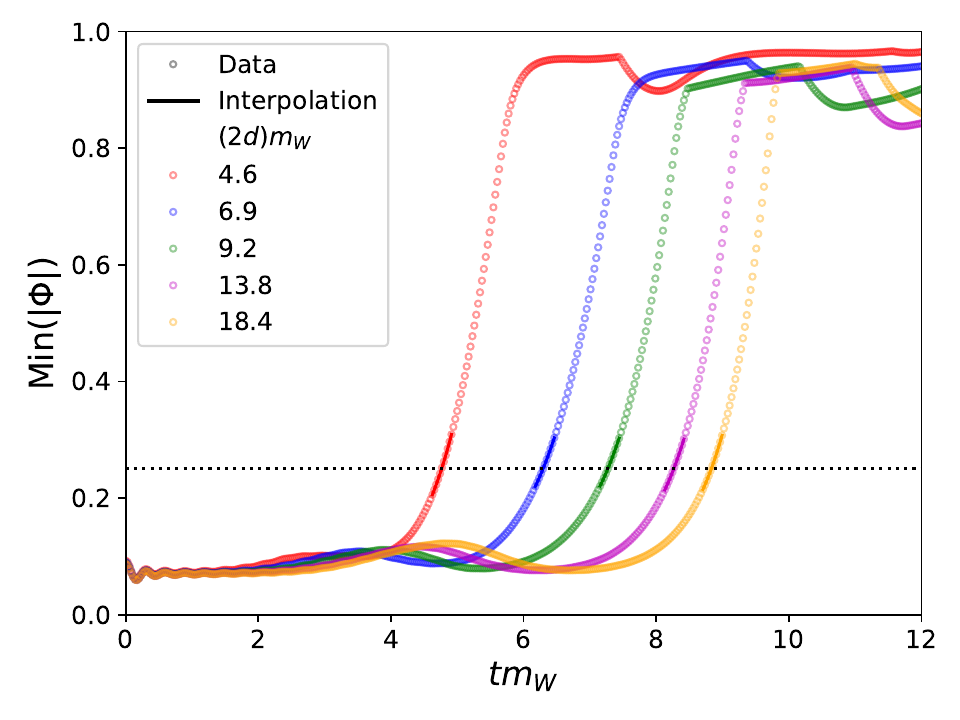}
\caption{The evolution of $\minphi$ in the simulation box for twists $\gamma=0$ (left panel)
and $\pi$ (right panel). The line colors correspond to different values of dumbbell lengths 
expressed as a multiple of the approximate monopole 
width $m_W^{-1}$. The hollow circles correspond to the data points from the simulation 
and the solid curves are $4^{\rm th}$ order polynomial fits around the threshold $\minphi=0.25$.
We show our threshold choice, $\minphi=0.25$, as a dotted horizontal line.}
\label{fig:minph-evo}
\end{figure}

The magnetic monopoles are zeros in the Higgs field and we leverage this to estimate 
the lifetime of the dumbbell by tracking the zeros and finding when they disappear in the 
simulation box.
Since the dumbbell is offset by $\delta/2$ in the positive $x$ and $y$ directions in our setup, 
the zeros of the Higgs field are never located at lattice points.
We instead borrow the approach 
from \cite{Vachaspati:2015ahr} used in studying the creation of monopoles via classical scattering.
We track the evolution of the minimum value of the Higgs field over the entire lattice, $\minphi$.
Once $\minphi$ exceeds a threshold, we tag the timestep in our simulation as the dumbbell 
lifetime.
The lifetime obtained by this criterion depends on the chosen value of the threshold.
As we will see, the result is sensitive to the threshold in the untwisted case but is
quite insensitive in the twisted case.
Additionally, we have tested the dependence of lifetimes on the spatial and time resolution of the 
simulation and demonstrated that there were no significant dependencies on the choice of our 
simulation parameters.

\begin{figure}
\includegraphics[width=0.5 \textwidth]{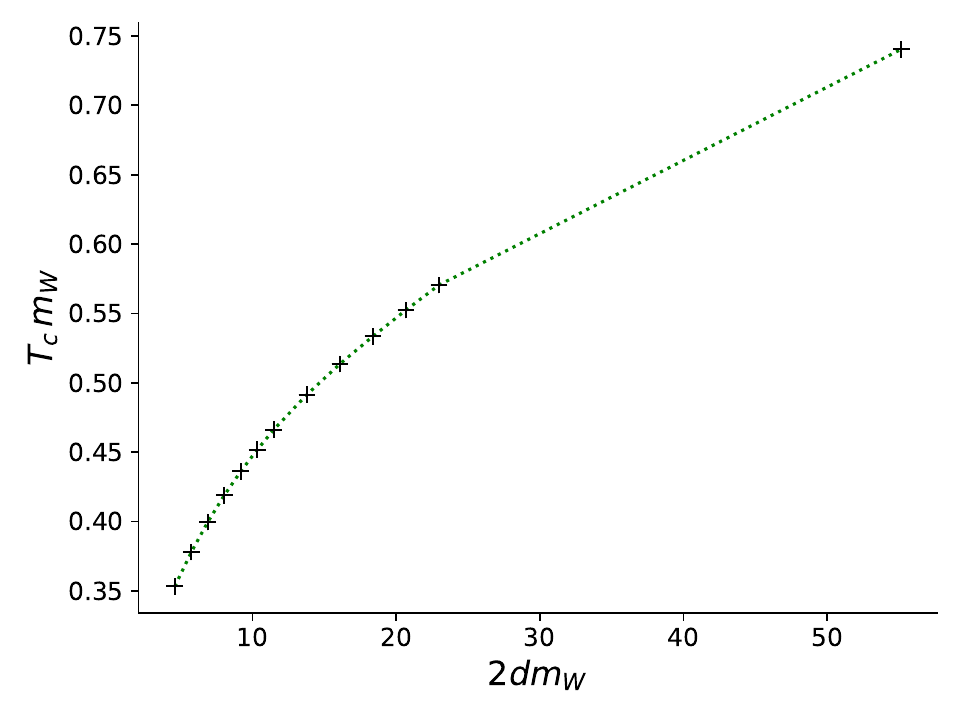}		
\includegraphics[width=0.5 \textwidth]{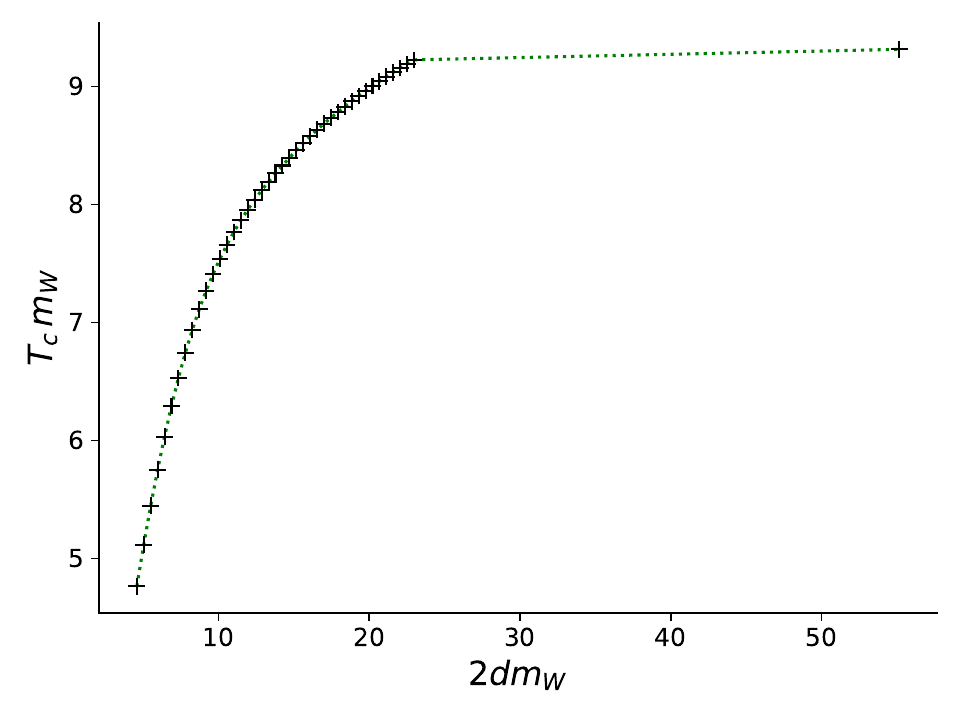}
\caption{The lifetimes $T_c$ of the dumbbells as a function of their initial separation are shown
for twists $\gamma=0$ and $\pi$ in the left and right panels, respectively.
}
\label{fig:Tc-2d_dependence}
\end{figure}

In Figure~\ref{fig:minph-evo}, we show the evolution of $\minphi$ for various values of initial 
separation $2d$. Comparing the untwisted (left panel) and twisted (right panel)
cases, respectively, we see that the curves for the untwisted case rise slowly, implying
greater sensitivity of the dumbbell lifetime to the chosen threshold. In the twisted case,
the curves rise very sharply and the dumbbell lifetime is not sensitive to our choice of
threshold.
After tagging the timestep when the threshold condition on $\minphi$ is satisfied in our simulations,
we interpolate the time evolution of $\minphi$ in a small range around the tagged timestep. 
The interpolation was performed via a fourth order polynomial curve fit, and
we evaluate the interpolated function at the threshold value ($\minphi=0.25$) to find the lifetime.
The interpolated $\minphi$ are shown in Figure~\ref{fig:minph-evo} as solid lines.

We plot the dumbbell lifetime against the initial dumbbell length for
untwisted and maximally twisted dumbbells in Figure~\ref{fig:Tc-2d_dependence}. We find
that the lifetime grows with initial length but appears to saturate beyond a certain length.
To check the saturation we have performed one run for very long dumbbells
($2dm_W=55.2$). These runs are computationally very expensive. The saturation is
clearer in the case of twisted dumbbells. We interpret these plots recalling the
instability analysis of Z-strings~\cite{James:1992wb,Goodband:1995he}: the untwisted 
dumbbell decays on a
timescale $\sim m_W^{-1}$ due to the instability, while the twisted dumbbell
survives about 10 times longer, first collapsing to an electroweak sphaleron that eventually 
decays due to its instability on a longer timescale.
To verify that the instability is dynamical and not a result of numerical artifacts,
we followed \cite{Achucarro:1999it,Urrestilla:2001dd} and ran test simulations with large 
values of $\sin^2\theta_w$ and small values of $m_H/m_Z$
for which Z-strings are known to be stable~\cite{James:1992wb,Goodband:1995he}. 
The results of these simulations are given
in Appendix~\ref{appnd:semilocal-dumbbell-sims}. In contrast to the simulations with
physical parameters, the Z-string instability is absent and the dynamics of the monopoles 
is clearly visible in these ``semilocal'' simulations.
	
\subsection{Relic Magnetic Fields}
\label{subsec:magneticfield}

\begin{figure}	
\includegraphics[width=0.5 \textwidth]{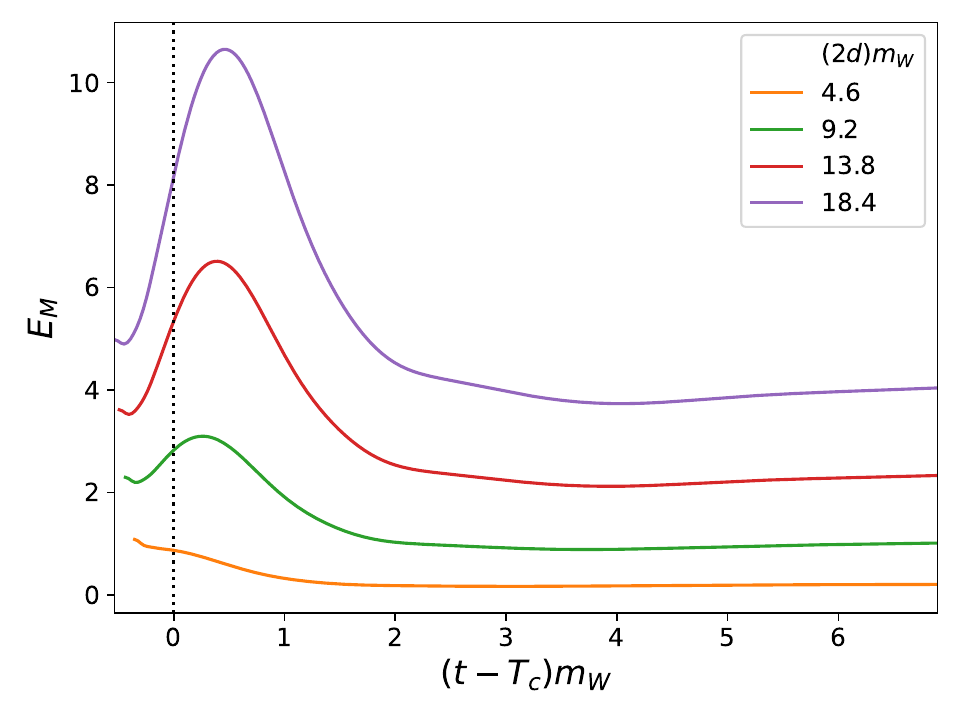}		
\includegraphics[width=0.5 \textwidth]{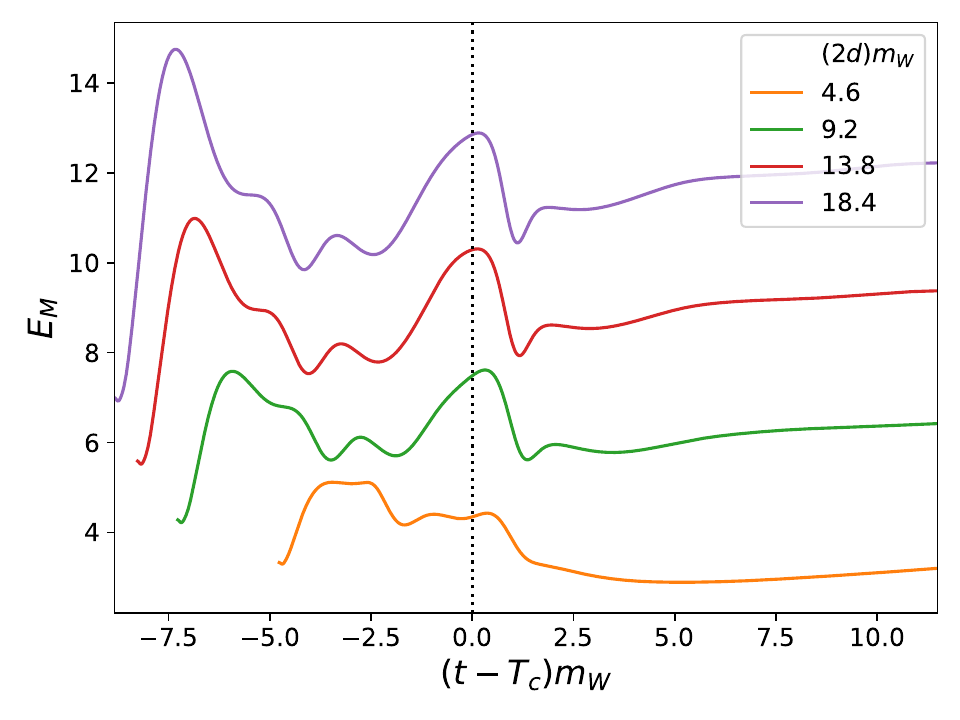}
\caption{
The total magnetic energy as a function of time for twists $\gamma=0$ and $\pi$ 
in the left and right panels, respectively. The time along the horizontal axis is shifted by the lifetime 
$T_c$ and $t=T_c$ is shown by a vertical dotted line.
Different colored curves correspond to different initial dumbbell lengths, as stated in the legends.
}
\label{fig:ME-t-dep}
\end{figure}
	
The definition for the electromagnetic field strength tensor in the
symmetry broken phase ($|\Phi|=\eta$) that
reduces to the usual Maxwell definition in unitary gauge is~\cite{tHooft:1974kcl,Vachaspati:1991nm},
\ba
\label{A field def proper}
A_{\mu\nu} &\equiv& - \sin\theta_w n^a W^a_{\mu\nu} + \cos \theta_w Y_{\mu\nu }
-i\frac{2\sin\theta_w}{g\eta^2}(D_\mu \Phi^\dagger D_\nu \Phi - D_\nu \Phi^\dagger D_\mu \Phi)\nn \\
&=&\partial_\mu A_\nu-\partial_\nu A_\mu
-i\frac{2\sin\theta_w}{g\eta^2}(\partial_\mu \Phi^\dagger \partial_\nu \Phi 
- \partial_\nu \Phi^\dagger \partial_\mu \Phi)\,.
\label{Amunu}
\ea
This definition implies the presence of non-zero electromagnetic fields even for $A_\mu=0$ 
due to the Higgs gradient term. 
In our study of the decay of dumbbells, $\partial_\mu\Phi$ vanishes at late times
and then the expression in \eqref{A field def proper} agrees with the Maxwell definition.

\begin{figure}
\begin{center}
\includegraphics[width=0.6 \textwidth]{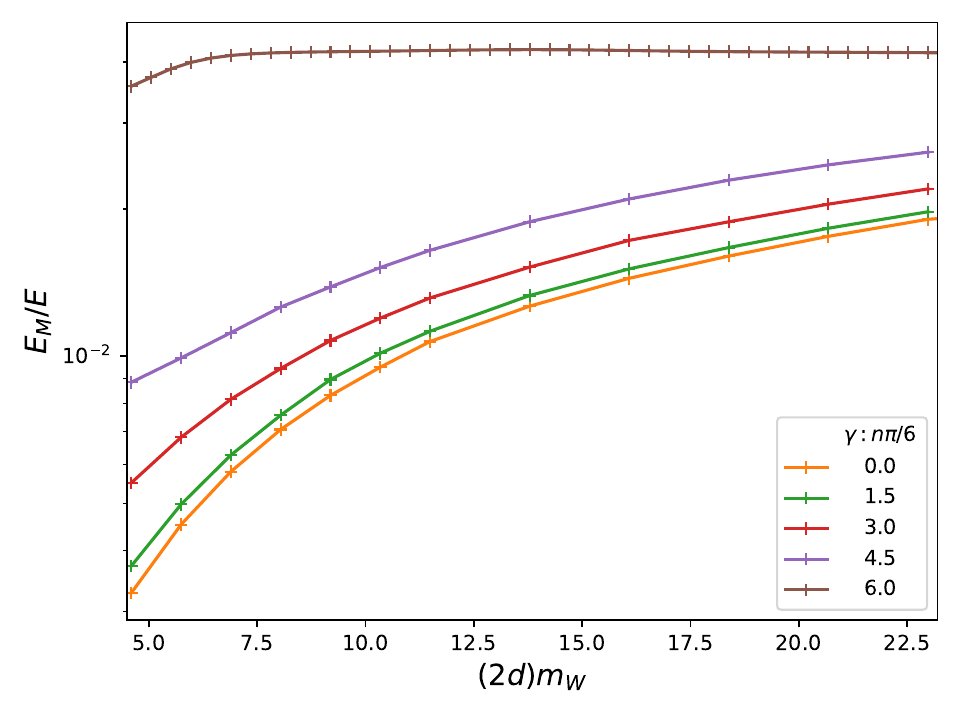}
\end{center}
\caption{
The magnetic energy at a late time, $E_M(T_c+25)$, as a function of 
initial dumbbell length. The different colors correspond to different values of twists 
$\gamma$ as stated in the legend.}
\label{fig:EMTc}
\end{figure}

The total magnetic field energy is given by
\begin{equation}\label{B-total-energy}
E_M(t)=\frac{1}{2}\int d^3x \, B^2\,.	
\end{equation}
We show the time evolution of magnetic energy $E_M(t)$ for twisted and untwisted dumbbells 
of various initial lengths in Figure~\ref{fig:ME-t-dep}.
After an initial phase of annihilation, the total magnetic field energy reaches a steady
value that depends on the initial length of the dumbbell.
An important observation is that the relic magnetic 
energy depends on the twist, in addition to the initial dumbbell lengths.
In Figure~\ref{fig:EMTc} we plot the fractional magnetic field energy at a late time after annihilation 
$T_c+\Delta t$, where we have 
chosen $\Delta t=25\eta^{-1}$. 
Since we use Dirichlet boundary conditions, the simulation time has a upper bound,
after which reflections occurring at the boundaries would affect quantities of interest.
The time at which we evaluate the relic magnetic field energy, $T_c+\Delta t$ is large enough 
such that the relic magnetic energy has reached an asymptote but is still smaller than the time 
at which a significant fraction of the energy is reflected.
From Figure~\ref{fig:EMTc}, we infer the fractional magnetic energy, after annihilation, is 
about twice as large
for the twisted case when compared to the untwisted case, for
the same initial (large) dumbbell length.

In addition to the magnetic energy density, we are interested in the helicity of the magnetic 
field, which has been shown to have significant implications for cosmic baryogenesis and 
magnetogenesis~\cite{Vachaspati_2001,VachaspaticandField,Vachaspati:2020blt}.
The total magnetic helicity is defined as
\begin{equation}\label{mag-helicity-def}
H_M(t)=\int d^3x \, \bfA\cdot \bfB\, ,
\end{equation}
where $\bfA$ is given by the spatial components of the vector potential 
\eqref{A def unitary gauge} and $\bfB$ is derived from the EM tensor \eqref{A field def proper}.

\begin{figure}
\begin{center}
\includegraphics[width=0.49 \textwidth]{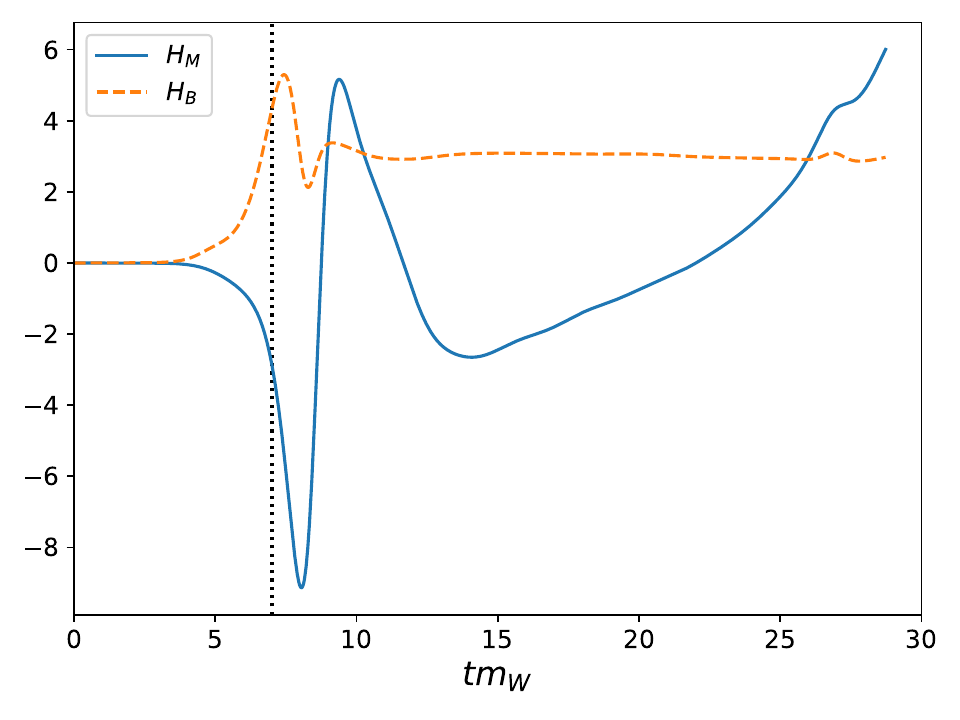}
\includegraphics[width=0.49 \textwidth]{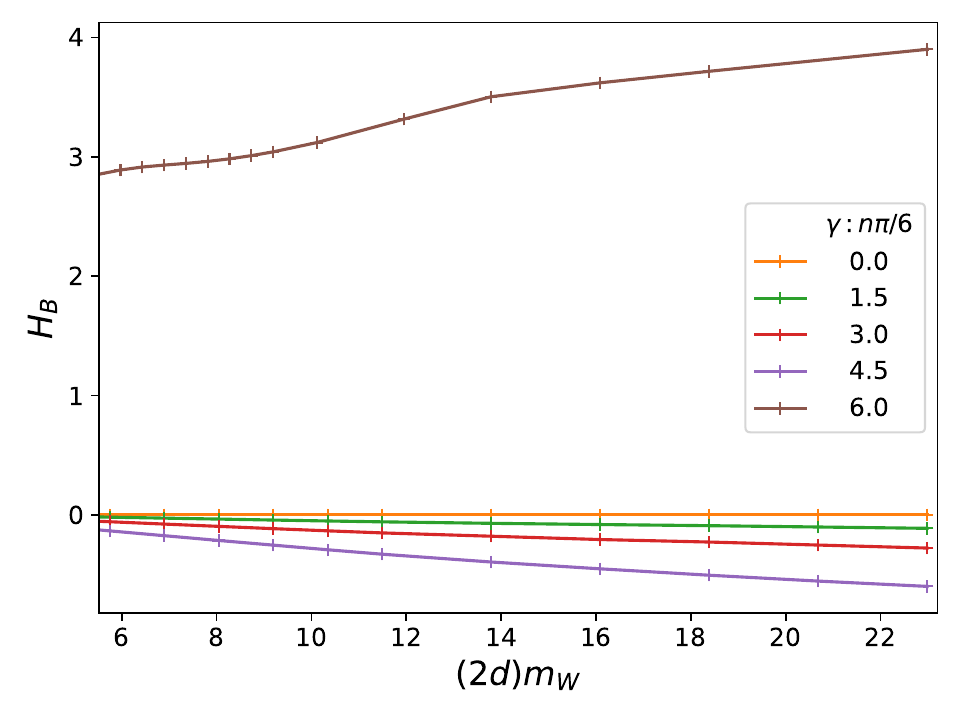}
\end{center}
\caption{
In the left panel, the magnetic helicity $H_M(t)$ (blue curve), and the physical helicity $H_B(t)$ (orange dashed curve) are shown
for the twisted case $\gamma=\pi$, with initial dumbbell length $2d = 80\delta$.
Here, the annihilation time $T_c$ is marked with a black dotted vertical line.
In the right panel, we plot the physical helicity at late time, $H_B(T_c+25\eta^{-1})$, as a function of the
initial dumbbell length, with different colors corresponding to the twist values shown in the legend.
}
\label{fig:hel-pi}
\end{figure}

The evolution of total magnetic helicity for an initially twisted dumbbell is shown by the blue 
curve in the left panel of Figure~\ref{fig:hel-pi}.
It is clear that the helicity does not approach a specific value within the span of our simulation.
A possible explanation for the behavior is that the definition of $H_M$  is gauge independent 
only if the magnetic field is
orthogonal to the areal vectors everywhere on the boundaries of the simulation domain.
(Alternately, if the magnetic field vanishes on the boundaries.)
This is certainly not true in our simulations. Hence we cannot assign
physical meaning to our evaluation of $H_M$ at late times when the magnetic field is not
small at the boundaries. 
Nonetheless, we can infer that the relic magnetic field has significant net helicity 
when compared to the untwisted case, which had a helicity $H\sim 10^{-13}$, consistent
with numerical roundoff.

\begin{figure}
\centering
\includegraphics[width=0.49\textwidth]{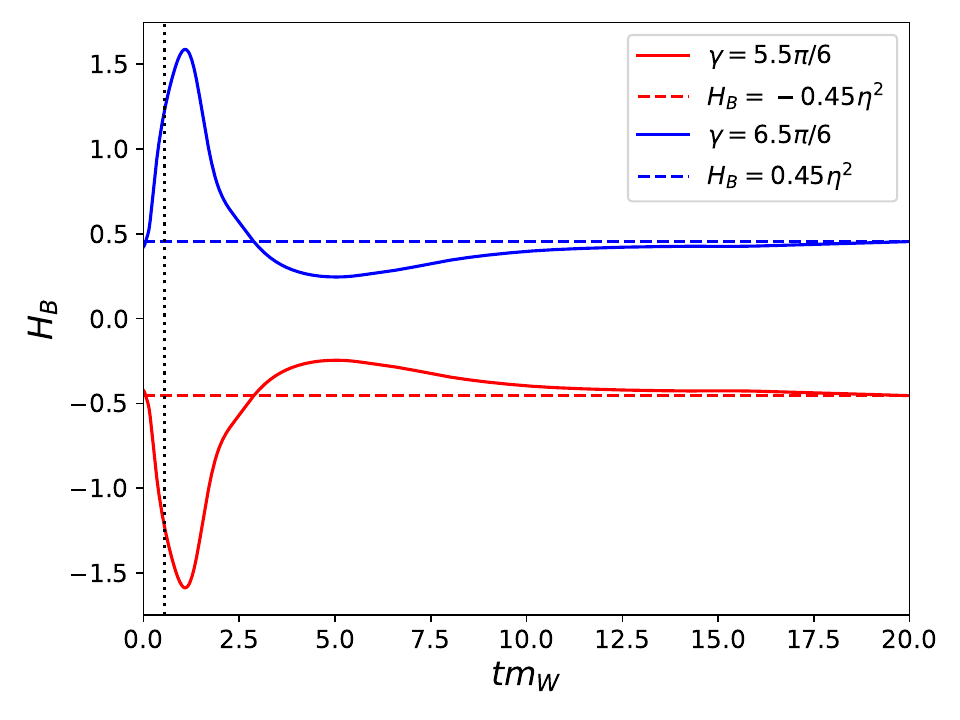}
\includegraphics[width=0.49\textwidth]{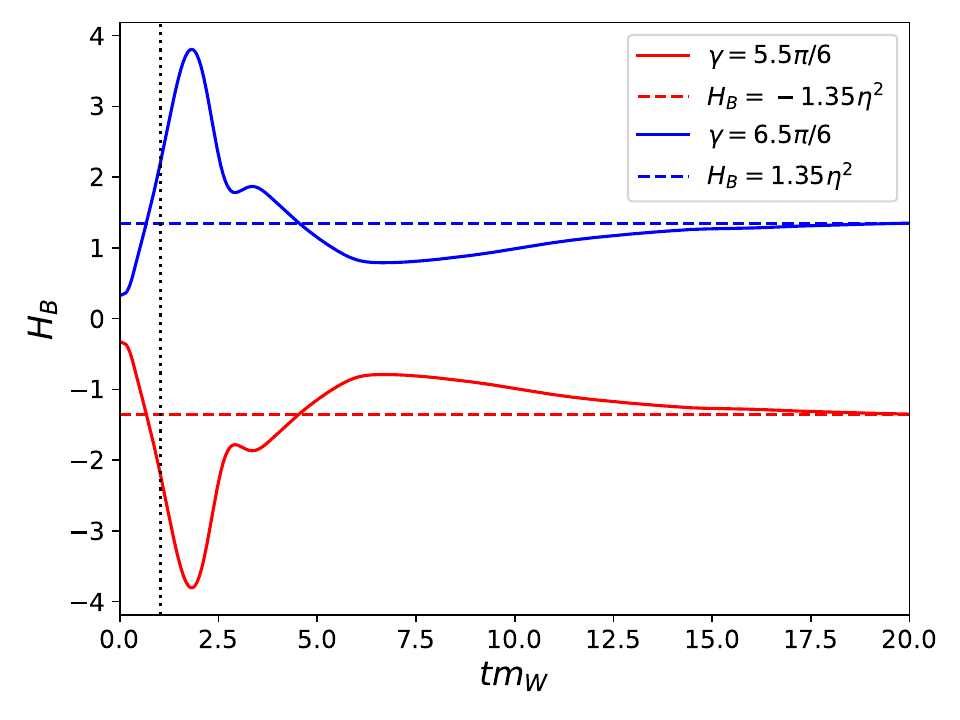}
\caption{
The evolution of physical helicity $H_B$ for twists $\gamma=5.5\pi/6$ (red curve) and $6.5\pi/6$ (blue curve) are shown for initial dumbbell lengths $2d=80\delta$ and $160\delta$ in the left and right panels, respectively.
The annihilation time $T_c$ is marked with a black dotted vertical line.
The horizontal red and blue curves correspond to the physical helicities at time $t=20m_W^{-1}$.
Here, the time step size in the simulation is $dt=\delta/8$, with $\delta=0.25\eta^{-1}$.
}
\label{fig:hel-comapare-gamma-pm}
\end{figure}

An alternative measure of the parity violation in the magnetic
field is provided by the ``physical helicity'' defined as,
\begin{equation}
H_B = \int d^3x \, \bfB \cdot (\nabla\times\bfB)\, .
\label{phs-hel}
\end{equation}
In contrast to the magnetic helicity, the physical helicity does not have issues 
with gauge invariance. It is also better behaved in a finite domain as the magnetic
field $\bfB$ falls off faster than the gauge field $\bfA$.
The evolution of $H_B$ is shown in Figure~\ref{fig:hel-pi} for $\gamma = \pi$,
and it asymptotes to a constant value at late times.
We also plot $H_B$ at late times after annihilation for various initial 
dumbbell lengths and twists in the right panel of Figure~\ref{fig:hel-pi}.
From this, we can infer that the $H_B$ for the untwisted case $\gamma=0$ is consistent with
zero. 
In the twisted case with $\gamma=\pi$, it takes on finite positive values, 
$\sim 3\eta^2$, as shown in Figure~\ref{fig:hel-pi}.
One feature, that is immediately clear, is that the signs of the physical helicities, $H_B$, for 
$\gamma=\pi$ and $0<\gamma<\pi$ are opposite to each other.
The opposite signs are related to the untwisting dynamics of the dumbbell configuration
as it annihilates. Dumbbells with twist $\gamma=\pi$ are maximally twisted and carry
maximal energy as a function of twist~\cite{Patel:2023sfm}. 
Dumbbells with twist slightly less than $\pi$ will tend
to untwist by reducing the twist angle from $\pi - \to 0$, while those with twist somewhat
greater than $\pi$ will untwist by increasing the twist angle from $\pi + \to 2\pi$. These
decay modes lead to magnetic field helicity of opposite signs.
We tested this argument by running the simulation for twists slightly larger and smaller than 
$\pi$, and confirmed that the signs of the physical helicity are indeed opposite to each other.
The plots for the tests we conducted, for two different initial dumbbell lengths, are shown in 
Figure~\ref{fig:hel-comapare-gamma-pm}, where we see that the evolution of $H_B$ for twist 
$\pi +$ mirrors the one for twist $\pi -$.

\section{Discussion and Conclusion}
\label{conclusions}	

We have studied the collapse, annihilation, and decay products of electroweak
dumbbells as a function of their length and twist. 

The untwisted case has the expected dynamics, where the dumbbell collapses in a time 
that is very short, $\sim m_W^{-1}$, and comparable to the instability time scale
of the Z-string~\cite{James:1992wb,Goodband:1995he}.
The energy of the dumbbell that is converted into magnetic field
energy depends on the length of the dumbbell. For short untwisted dumbbells, the magnetic
field energy can be $\sim 0.1\%$; for longer dumbbells it is $\sim 1\%$. 
The magnetic fields produced in the untwisted case are not helical.

Twisted dumbbells collapse on a time-scale that saturates to $\sim 10 m_W^{-1}$ for
long dumbbells.
They collapse to form a long-lived object that presumably is
an electroweak sphaleron, see Figure~\ref{fig:energy-slices-twist-pi}.
Eventually, the sphaleron decays as well. The decay
products include magnetic field energy that is large compared to the untwisted
case. The fractional energy converted to magnetic field energy is roughly 
independent of the dumbbell length and is $\sim 4\%$. The produced magnetic
field is helical though our calculation of the magnetic helicity is not reliable,
especially at late times, due to the finiteness of the simulation box. As an
alternative we have calculated the integrated physical helicity and this asymptotes 
to a constant ($\sim 3\eta^2$), 
at late times for the maximally twisted dumbbell.

In future work, we propose to study the dynamics of rotating dumbbells as
this pertains to their production and detection in a laboratory setting as first
discussed by Nambu \cite{Nambu:1977ag}. We expect the study to be technically
challenging as it would require significant improvements in implementing
boundary conditions, especially if rotating dumbbells survive for a long time.

\acknowledgments

This work was supported by the U.S. Department of Energy, Office of High Energy 
Physics, under Award No.~DE-SC0019470.
The authors acknowledge Research Computing at Arizona State University for providing access to high performance computing and storage resources on the Sol Supercomputer that have contributed to the research results reported within this paper.

\bibliographystyle{JHEP}
\bibliography{paper}

\newpage
	
\appendix

\section{Semilocal Dumbbell simulations}
\label{appnd:semilocal-dumbbell-sims}
\begin{figure}[H]
\includegraphics[width=\textwidth]{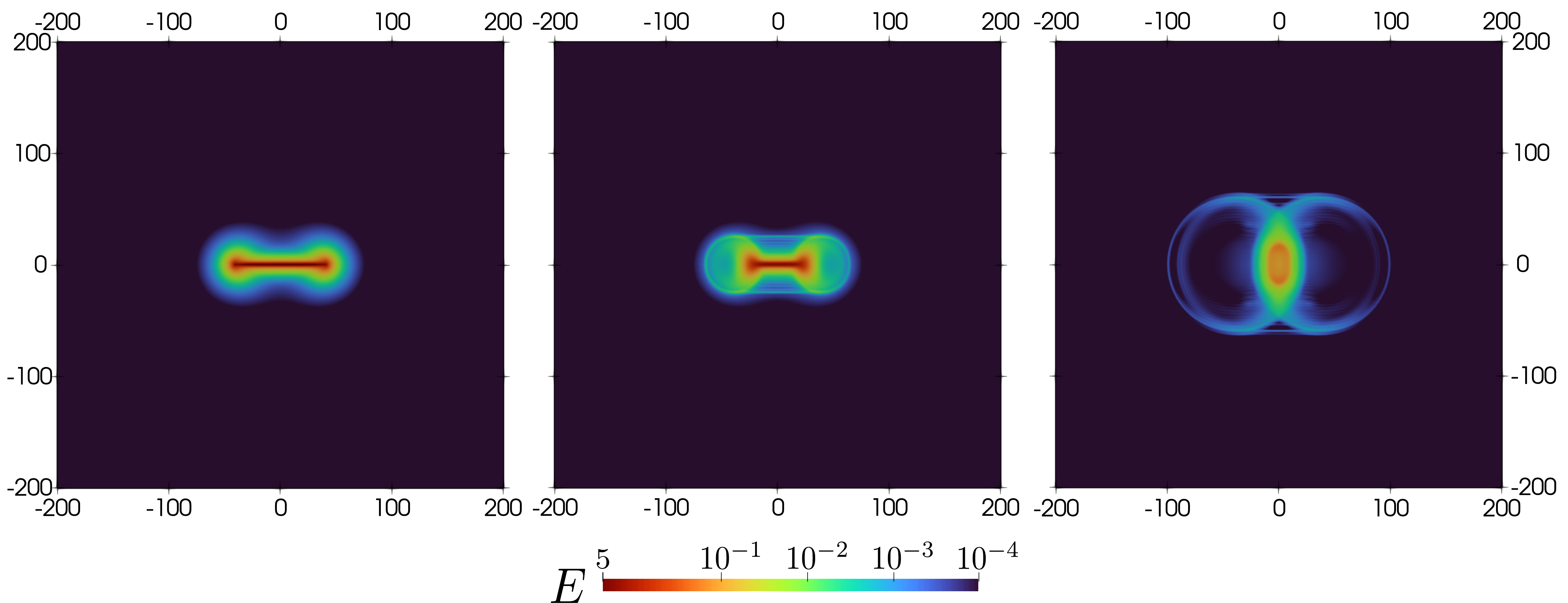}
\caption{
The snapshots of energy density, for dumbbell evolution in the semilocal limit, in $xz$-plane and twist $\gamma=0$ are shown here. 
The left, middle and right panels correspond to times $0dt$, $200dt$ and $500dt$, respectively.
The colors represent the energy density and the corresponding values are given by the scale, in units of $\eta^4$.
Here, the simulation parameters are $dt=\delta/8$, with $\delta=0.25\eta^{-1}$, and the initial dumbbell length is $2d=80\delta$.
The horizontal ($z$) and vertical ($x$) axis are in lattice units.
}
\label{fig:semilocal-energy-slices-tw-0}
\end{figure}

\begin{figure}
\includegraphics[width=\textwidth]{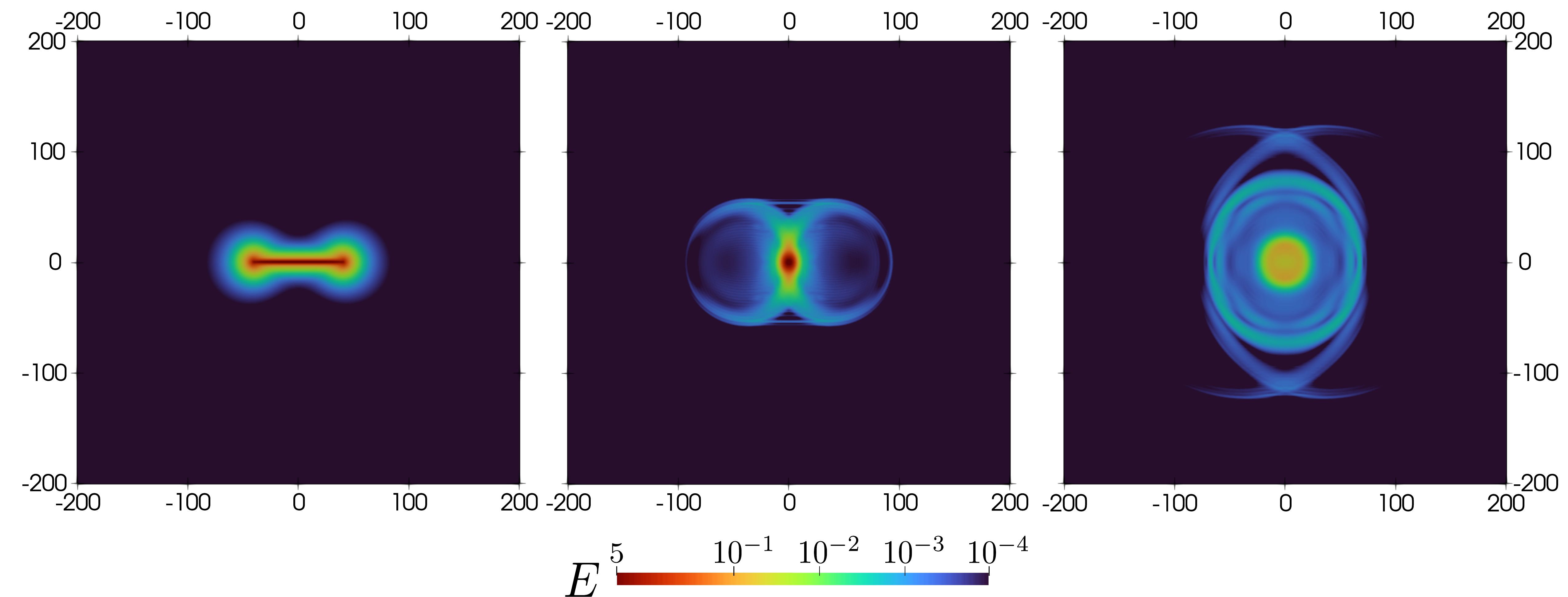}
\caption{
The energy density snapshots, similar to Figure~\ref{fig:energy-slices-twist-0}, for the twisted case ($\gamma=\pi$) and times $0dt$, $450dt$ and $1000dt$ in the left, middle and right panels, respectively.
}
\label{fig:semilocal-energy-slices-tw-pi}
\end{figure}

\begin{figure}
\centering	
\includegraphics[width=0.5\textwidth]{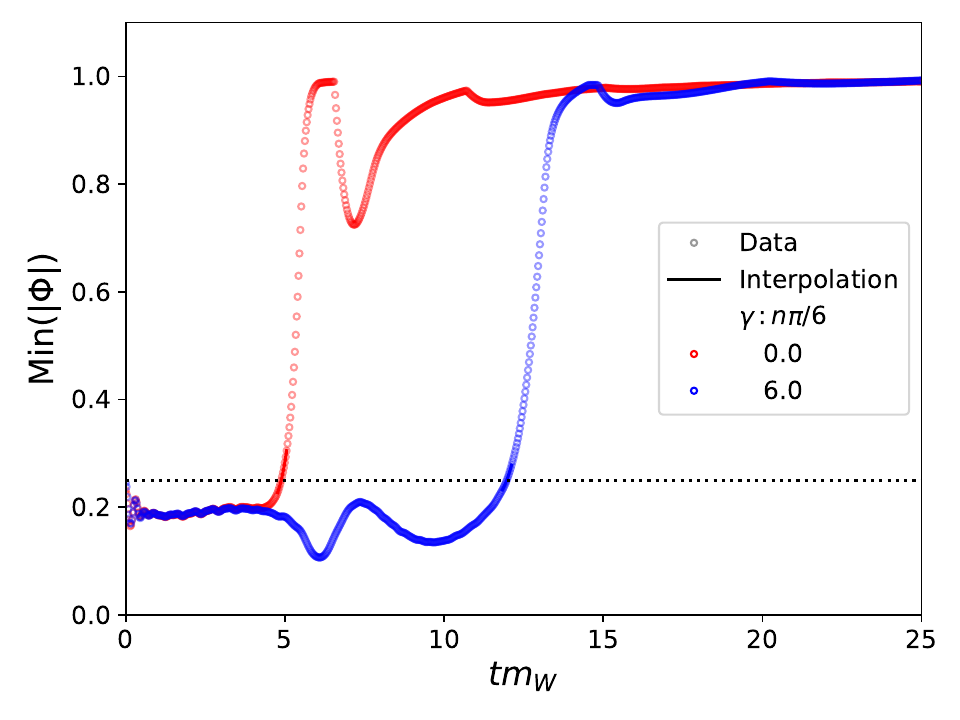}
\caption{The evolution of $\minphi$ in the simulation box for twists $\gamma=0$ (red curve)
and $\pi$ (blue curve). The hollow circles correspond to the data points from the simulation 
and the solid lines are $4^{\rm th}$ order polynomial fits around the threshold $\minphi=0.25$.
We show our threshold choice, $\minphi=0.25$, as a dotted horizontal line.}
\label{fig:minph-evo-semilocal}
\end{figure}

The Z-string has been shown to be stable in the semilocal limit \cite{Achucarro:1999it}, which 
corresponds to $\sin^2\theta_w=1$ and $\beta \equiv m_H^2 / m_Z^2 < 1$.
As a test of our simulation code, we ran an instance of the dumbbell simulation for the parameters 
$\sin^2\theta_w=0.995$ and $\beta=0.01$, which lie in the stable regime stated 
in~\cite{Achucarro:1999it} and studied for a cosmological distribution of dumbbells 
in~\cite{Urrestilla:2001dd}.
We show several snapshots of the energy density in the $xz$-plane for the untwisted 
case ($\gamma=0$) in Figure~\ref{fig:semilocal-energy-slices-tw-0}, and for the twisted case 
($\gamma=\pi$) 
in Figure~\ref{fig:semilocal-energy-slices-tw-pi}.
Unlike the electroweak case, shown in Figures~\ref{fig:energy-slices-twist-0}
and ~\ref{fig:energy-slices-twist-pi}, it can be seen that the Z-string is stable, and the 
monopole-antimonopole pair move towards each, eventually undergoing annihilation.
For the twisted case, we once again observe the formation of an intermediate stable object 
which eventually decays, and can be seen in Figure~\ref{fig:semilocal-energy-slices-tw-pi}.
The evolution of $\minphi$ are shown in Figure~\ref{fig:minph-evo-semilocal}, and the lifetimes 
are $T_cm_W = 4.9$ and $12.0$, for the untwisted ($\gamma=0$) and twisted ($\pi$) cases, respectively.
These lifetimes are much longer than those of the dumbbell in the electroweak case for the same lengths; $T_cm_W = 0.4$ and $7.3$, for twists $\gamma=0$ and $\pi$, respectively.

\end{document}